\documentclass[aps,
prb,
preprint,
superscriptaddress,
longbibliography]{revtex4-2}

\usepackage{graphicx}
\usepackage{multirow}
\usepackage{amsthm,amsmath,amsfonts}
\usepackage{comment}
\usepackage{hyperref}
\usepackage{longtable,tabularx}
\usepackage{subfigure,xcolor,color}
\usepackage{siunitx}
\usepackage[utf8]{inputenc}
\usepackage[ruled,vlined]{algorithm2e}
\usepackage{graphicx}
\usepackage{dcolumn}
\usepackage{float}
\usepackage{bbm,url}
\usepackage{tabularx}
\usepackage{bm}
\usepackage{amssymb}
\usepackage{mathtools}

\begin{document}

\title{Modal Backflow Neural Quantum States for Anharmonic Vibrational Calculations
}

\author{Lexin Ding}
\email{lexin.ding@phys.chem.ethz.ch}
\affiliation{ETH Z\"urich, Department of Chemistry and Applied Biosciences, Vladimir-Prelog-Weg 2, CH-8093 Z\"urich, Switzerland}

\author{Markus Reiher}
\email{mreiher@ethz.ch}
\affiliation{ETH Z\"urich, Department of Chemistry and Applied Biosciences, Vladimir-Prelog-Weg 2, CH-8093 Z\"urich, Switzerland}

\begin{abstract}
    Neural quantum states (NQS) are a promising ansatz for solving many-body quantum problems due to their inherent expressiveness. Yet, this expressiveness can only be harnessed efficiently for treating identical particles if the suitable physical knowledge is hardwired into the neural network itself.
    For electronic structure, NQS based on backflow determinants has been shown to be a powerful ansatz for capturing strong correlation. By contrast, the analogue for bosons, backflow permanents, is unpractical due to the steep cost of computing the matrix permanent and due to the lack of particle conservation in common bosonic problems.
    To circumvent these obstacles, we introduce a modal backflow (MBF) NQS design and demonstrate its efficacy by solving the anharmonic vibrational problem. To accommodate the demand of high accuracy in spectroscopic calculations, we implement a selected-configuration scheme for evaluating physical observables and gradients, replacing the standard stochastic approach based on Monte Carlo sampling. A vibrational self-consistent field calculation is conveniently carried out within the MBF network, which serves as a pretraining step to accelerate and stabilize the optimization. In applications to both artificial and \textit{ab initio} Hamiltonians, we find that the MBF network is capable of delivering spectroscopically accurate zero-point energies and vibrational transitions in all anharmonic regimes. 
\end{abstract}

\maketitle

\clearpage

\section{Introduction}
A central endeavor in quantum chemistry is the efficient solution of the many-body Schr\"odinger equation. In electronic structure theory, this involves solving the electronic Hamiltonian with the Coulomb interaction in an exponentially large Hilbert space. Since exact diagonalization is only feasible for around two dozens of half-filled spatial orbitals,\cite{vogiatzis2017pushing,gao2024distributed} tailored wavefunction ans\"atze and optimization schemes have been developed to bypass the exponential scaling. Examples are 
selected configuration interaction (see, e.g., \cite{bender1969studies,schriber2016communication,holmes2016heat,liu2016ici,zimmerman2017incremental,chilkuri2021comparison}),
tensor network states 
(TNSs),\cite{white1992,baiardi2020density,ma2022density} 
full configuration interaction quantum Monte Carlo,\cite{booth2009fermion,eriksen2020shape} and auxiliary field quantum Monte Carlo.\cite{zhang2003afqmc,lee2022twenty}  
Despite the tremendous improvements in the past few decades, common wavefunction methods still suffer from at least one of the following shortcomings: The failure to include both static and dynamical correlation, steep scaling in the number of parameters to be optimized for strongly correlated systems, and the lack of efficient optimization methods.

In an effort to overcome these shortcomings, Carleo and Troyer proposed a novel type of quantum state ansatz using artificial neural networks (ANN).\cite{carleo2017solving} Their idea was to leverage the expressiveness and flexible designs of ANNs,\cite{rumelhart1986learning,lecun1989backpropagation,specht1991general} as well as the knowledge on their optimization.\cite{robbins1951stochastic,polyak1964some,duchi2011adaptive,kingma2014adam} After all, a quantum state is nothing but a function that maps particle positions to complex numbers, which, according to the universal approximation theorem, is formally learnable to arbitrary accuracy by feedforward neural networks.\cite{cybenko1989approximation,hornik1989multilayer} 
We note that deep feedforward neural networks are more expressive than tensor networks\cite{sharir2022neural} and that neural networks can even simulate volume-law entanglement with only polynomially many degrees of freedom.\cite{deng2017entanglement,zakari2025comment} These seminal results quickly ignited a new type of approximation, now known as neural quantum states (NQSs).\cite{luo2019backflow,manzhos2020machine,hermann2020deep,pfau2020ab,robledo2022fermionic,scherbela2022solving,hermann2023ab,zhao2023scalable} For the electronic ground state problem in small molecules, NQSs can even reach the accuracy obtained with
full configuration interaction.
\cite{pfau2020ab,choo2020fermionic,yoshioka2021solving}

Compared to the large body of work on electronic NQSs, application to the bosonic vibrational problem is somewhat less explored and limited to the first-quantized form.\cite{zhang2024neural} 
In contrast to lattice Hamiltonians, 
the vibrational problem has much less structure, in that every pair of vibrational modes is connected by the Hamiltonian.
In analogy to correlated wavefunction ans\"atze in electronic structure theory, there exist vibrational counterparts, including
vibrational configuration interaction methods\cite{bowman1979application,garnier2016adaptive,schroder2021incremental,schroder2022vibrational} and vibrational coupled cluster theories.\cite{christiansen2004vibrational,faucheaux2015higher,madsen2018tensor,christiansen2022}
Recently, the vibrational density matrix renormalization group (vDMRG) optimizing a TNS ansatz has become an alternative for calculating accurate molecular spectra.\cite{baiardi2017vibrational,baiardi2019optimization,larsson2019computing,glaser2022,larsson2025benchmarking}
However, matrix product states (MPSs) and tree tensor network states (TTNSs) assume certain topologies of the correlation of the vibrational modes, which conceptually clashes with the structureless character of the vibrational Hamiltonian.
By contrast, fully connected neural networks assume no such topologies.

Under the Born-Oppenheimer approximation, 
the vibrational Hamiltonian is comprised by a kinetic term and by a representation of the potential energy surface (PES, given by the electronic energy at different nuclear positions), which determines the achievable accuracy crucially. At first sight, the vibrational problem seems easier than the electronic one, due to the absence of long-range Coulomb interactions. 
However, for a solution to be predictive for vibrational spectroscopy, one must be able to resolve closely lying excited states while dealing with highly many-body coupling terms (typically up to six-body in practice, whereas the Coulomb interaction is only two-body) in the anharmonic vibrational Hamiltonian.

Although for a moderate number of normal modes (about two dozens) the spectroscopic accuracy of 1 cm$^{-1}$ \cite{puzzarini2019accuracy} is routinely reachable with multilayer multiconfigurational time-dependent Hartree \cite{wang2003multilayer,manthe2008multilayer,vendrell2011multilayer} and TNSs methods \cite{baiardi2017vibrational,baiardi2019optimization,larsson2019computing,glaser2023flexible,larsson2025benchmarking}, 
a larger number of modes can only be handled for lattice Hamiltonians
with a limited connectivity of interactions. \cite{vendrell2011multilayer} Hence, in view of the properties of NQSs, it is natural to explore and assess their capabilities for the vibrational structure problem. We note that the solution of the vibrational Hamiltonian is a fundamental building block for more complex problems in quantum chemistry, such as vibronic effects \cite{fulton1961vibronic,meng2013multilayer} or vibrational dynamics inside cavities.\cite{wang2021roadmap,dunkelberger2022vibration}

That being said, the universal approximation theorem \cite{cybenko1989approximation,hornik1989multilayer}
alone does not guarantee that any neural network can be efficiently trained to solve a given problem. Successful implementations of NQSs for the electronic Hamiltonians showed that it is crucial to incorporate physical knowledge, such as the antisymmetrization of electronic wavefunctions and cusp conditions, into the networks.\cite{pfau2020ab} Another important insight from the earlier work is that using ANN to learn indirect features, such as Jastrow factors and backflow determinants, is more efficient than directly learning the wavefunction.\cite{luo2019backflow,hermann2020deep,liu2024backflow,zakari2025comment} In particular, backflow determinants (Slater determinants computed from backflow transformed orbitals) plays a crucial role in compressing exponential degrees of freedom in highly entangled fermionic systems.\cite{passetti2023can,zakari2025comment} The starting point of this work is therefore to identify the physical knowledge in the bosonic vibrational problem that can be incorporated into the network design. Specifically, we set out to develop the bosonic analog of backflow transformed orbitals, which turns out to be not the vibrational modes, but rather basis functions of the single-mode Fock spaces, i.e., modals. 
Based on this insight, we propose the modal backflow (MBF) NQSs and demonstrate their capability of solving the vibrational problem at different degrees of anharmonicity. 

This paper is structured as follows. In Section \ref{sec:ham} we briefly review the Watson Hamiltonian in second-quantized form, which will serve as our vibrational model Hamiltonian here. In Section \ref{sec:mbf}, we introduce the MBF NQSs as an ansatz to solve the vibrational problem, at the example of the Watson Hamiltonian. In Section \ref{sec:num}, we investigate MBF NQSs on a set of randomly generated Watson Hamiltonian at different levels of anharmonicity. And finally in Section \ref{sec:ab}, we apply the MBF  NQS wavefunction to ab initio vibrational Hamiltonians of molecules and solve for the zero-point energy and low-lying transition energies.

\section{Anharmonic Vibrational Hamiltonian} \label{sec:ham}

A general PES may be approximated by a finite-order Taylor expansion around the equilibrium structure of a molecule. In the normal coordinates $\hat{q}_i$ where the Hessian of the PES is diagonal, the Watson Hamiltonian with $L$ degrees of freedom takes the following form\cite{watson1968simplification}
\begin{equation}
\begin{split}
    H_{\rm vib} = &\frac{1}{2}\sum_{i=1}^L w_i (\hat{p}_i^2 + \hat{q}_i^2) 
    \\
    + &\frac{1}{6} \sum_{i,j,k=1}^{L} \Phi_{ijk}^{(3)} \hat{Q}_{ijk} + \frac{1}{24} \sum_{i,j,k,l=1}^{L} \Phi_{ijkl}^{(4)} \hat{Q}_{ijkl} + \cdots
    \\
    + &\sum_{i,j,k,l=1}^L \sum_{\tau=x,y,z} B^\tau \xi_{ij}^\tau \xi_{kl}^\tau \sqrt{\frac{w_jw_l}{w_iw_k}} \hat{q}_i\hat{p}_j \hat{q}_k \hat{p}_l.
\end{split}
\end{equation}
Here, $w_i$'s are harmonic frequencies of the normal modes, and we use the shorthand notation $\hat{Q}_{ijk} = \hat{q}_i\hat{q}_j \hat{q}_k$ and $\hat{Q}_{ijkl} = \hat{q}_i\hat{q}_j \hat{q}_k \hat{q}_l$. $\bm{\Phi}^{(3)}$ and $\bm{\Phi}^{(4)}$ are the third- and fourth-order reduced force constants, defined as
\begin{equation}
    \Phi_{ijk}^{(3)} = \frac{\kappa_{ijk}^{(3)}}{\sqrt{w_iw_jw_k}}, \quad \Phi_{ijkl}^{(4)} = \frac{\kappa_{ijkl}^{(4)}}{\sqrt{w_iw_jw_kw_l}},
\end{equation}
where $\bm{\kappa}^{(3)}$ and $\bm{\kappa}^{(4)}$ are the third-and fourth-order partial derivatives of the PES evaluated at the local minimum of the equilibrium structure, respectively. They provide anharmonic corrections to the harmonic approximation of the vibrational Hamiltonian. Additionally, the Watson Hamiltonian includes the Coriolis terms that couple the positions $\hat{q}_i$ and momenta $\hat{p}_i$ defined by the rotational constants $B^\tau$ and Coriolis coupling constants $\xi^\tau_{ij}$.\cite{carbonniere2004coriolis}

To arrive at the second-quantized form of the anharmonic Hamiltonian, we substitute the position and momentum operators with
\begin{equation}
    \hat{p}_i = \frac{1}{\sqrt{2}}(b^\dagger_i + b^{\phantom{\dagger}}_i), \quad \hat{q}_i = \frac{1}{\sqrt{2}}(b^\dagger_i - b^{\phantom{\dagger}}_i),
\end{equation}
where $b^{}_i$ $(b^{\dagger}_i)$ are bosonic annihilation (creation) operators of mode $i$ with actions on the occupation number vector (ONV) as follows
\begin{equation}
\begin{split}
    b^{\phantom{\dagger}}_i |n_1\cdots n_i \cdots n_L\rangle &= \begin{cases}
        \sqrt{n_i} |n_1\cdots n_i\!-\!1 \cdots n_L\rangle, \quad &n_i \!>\! 0
        \\
        0, \quad &n_i \!=\! 0
    \end{cases},
    \\
     b^\dagger_i |n_1\cdots n_i \cdots n_L\rangle &= \sqrt{n_i\!+\!1} |n_1\cdots n_i\!+\!1 \cdots n_L\rangle.
\end{split}
\end{equation}
The substitution above then allows us to rewrite the Hamiltonian in second-quantized form,
\begin{equation}
\begin{split}
    \hat{H}_{\rm vib} = &\sum_{i=1}^L w_i \left(b^\dagger_i b^{\phantom{\dagger}}_i + \frac{1}{2}\right) 
    \\
    &+ \frac{1}{12\sqrt{2}} \sum_{i,j,k=1}^{L} \Phi_{ijk}^{(3)} \prod_{r=i,j,k}\!(b^\dagger_r\! +\! b^{\phantom{\dagger}}_r) 
    \\
    &+ \frac{1}{96} \sum_{i,j,k,l=1}^{L} \Phi_{ijkl}^{(4)} \prod_{r=i,j,k,l}\!(b^\dagger_r \!+\! b^{\phantom{\dagger}}_r) + \cdots
    \\
    &+ \frac{1}{4} \sum_{i,j,k,l=1}^L \sum_{\tau=x,y,z} B^\tau \xi_{ij}^\tau \xi_{kl}^\tau \sqrt{\frac{w_jw_l}{w_iw_k}} 
    \\
    &\times  (b^\dagger_i\! +\! b^{\phantom{\dagger}}_i) (b^\dagger_j\! -\! b^{\phantom{\dagger}}_j) (b^\dagger_k\! +\! b^{\phantom{\dagger}}_k) (b^\dagger_l\! -\! b^{\phantom{\dagger}}_l). 
\end{split}
\end{equation}
The Hilbert space $\mathcal{H}$ on which $\hat{H}_{\rm vib}$ acts is, in principle, infinite.
In practice, however, we impose a cutoff $N_{\rm max}$ as the maximum number of particles in each mode, or more generally $N_{\rm modal}\!=\!N_{\rm max}\!+\!1$ as the number of basis states, also known as modal functions. Typically, $N_{\rm modal}$ is higher than the number of basis states in an electronic orbital, which is 4. This leads to a steeper exponential scaling $(N_{\rm modal})^L$ in the Hilbert space dimension in terms of the number of vibrational modes $L$, compared to $4^L$ in the electronic case, where $L$ now refers to the number of electronic orbitals.

The cutoff $N_{\rm max}$ for the mode occupancies (or $N_{\rm modal}$ for the number of modal basis functions) also leads to restrictions of bosonic operators to the truncated Hilbert space. We denote such restrictions as $[\cdot]_{N_{\rm max}}$. It is straightforward to write down the restricted annihilation (creation) operator $[b^{}_i]_{N_{\rm max}}$ ($[b^{\dagger}_i]_{N_{\rm max}}$)
\begin{equation}
    \begin{split}
        [b^{\phantom{\dagger}}_i]_{N_{\rm{max}}} &= \begin{pmatrix}
            0 & 0 & 0 & 0  & 0
            \\
            1 & 0 & 0  & 0 & 0
            \\
            0 & \sqrt{2} & 0 &  0 & 0
            \\
            \vdots & \vdots & \ddots & \vdots & 0 
            \\
            0 & 0 & 0 & \sqrt{N_{\rm max}} & 0 
        \end{pmatrix},
        \\
        [b^\dagger_i]_{N_{\rm{max}}} &= \begin{pmatrix}
            0 & 1 & 0 & \cdots  & 0
            \\
            0 & 0 & \sqrt{2}  & \cdots & 0
            \\
            0 & 0 & 0 &  \ddots & 0
            \\
            0 & 0 & 0 & \cdots & \sqrt{N_{\rm max}} 
            \\
            0 & 0 & 0 & 0 & 0 
        \end{pmatrix}.
\end{split}
\end{equation}
However, the restriction of strings of creation/annihilation operators, such as $\hat{N}_i = b^\dagger_{i}b^{\phantom{\dagger}}_i$, cannot be constructed by simply multiplying $[b^{\dagger}_i]_{N_{\rm max}}$ and $[b^{}_i]_{N_{\rm max}}$, as one can verify that
\begin{equation}
\begin{split}
        [b^\dagger_i]_{N_{\rm{max}}}[b^{\phantom{\dagger}}_i]_{N_{\rm{max}}} \neq [\hat{N}_i]_{N_{\rm{max}}} &= \begin{pmatrix}
            0 & 0 & 0 & \cdots & 0 
            \\
            0 & 1 & 0 & \cdots & 0
            \\
            0 & 0 & 2 & \cdots & 0
            \\
            \vdots & \vdots & \vdots & \ddots & \vdots
            \\
            0 & 0 & 0 & \cdots & N_{\rm max}
        \end{pmatrix}.
    \end{split}
\end{equation}
To consistently construct the restrictions of bosonic operator strings, $[b^{\dagger}_i]_{N'_{\rm max}}$ and $[b^{}_i]_{N'_{\rm max}}$ should be first constructed in a larger Hilbert space with $N'_{\rm max} > N_{\rm max}$, then multiply to form the bosonic operator string, and finally truncated to the desired dimension $N_{\rm max}$. The difference $\Delta N_{\rm max} = N'_{\rm max} - N_{\rm max}$ should be greater or equal to $l-1$ where $l$ is the length of the bosonic operator string. For example, in the case of $b^\dagger_i b^{\phantom{\dagger}}_i$ we have $l=2$, and the following relation holds
\begin{equation}
    \left[[b^\dagger_i]_{N_{\rm{max}}+1}[b^{\phantom{\dagger}}_i]_{N_{\rm{max}}+1}\right]_{N_{\rm max}} = [\hat{N}_i]_{N_{\rm{max}}}.
\end{equation}

\section{Modal Backflow Neural Quantum States}\label{sec:mbf}

\subsection{Network Ansatz}

The concept of a backflow transformation dates back to Feynmann and Cohen in the 1950s.\cite{feynman1956backflow} It adds a multi-particle component to the single-particle coordinates\cite{feynman1956backflow,kwon1993backflow} to introduce a reverse flow of particle current, hence the term backflow. 
As the concept of backflow becomes more general, multi-particle dependency can also be mediated using occupation numbers,\cite{tocchio2008backflow} while the backflow transformed quantities can also be single-particle functions (orbitals) 
instead of positions.\cite{luo2019backflow,liu2024backflow} In particular, neural backflow (NBF) ans\"atze for fermions have been shown to be significantly more efficient in capturing the correlation of indistinguishable fermions compared to a standard feedforward neural network (FNN).\cite{luo2019backflow,liu2024backflow,passetti2023can,zakari2025comment}
We therefore briefly recall the formulation of the NBF ansatz for $N$ spinless fermions. Let $\bm M$ be an $L\!\times\!N$ matrix,
\begin{equation}
    \bm{M} = \begin{pmatrix}
        \varphi_{1,1} & \varphi_{2,1} & \cdots & \varphi_{N,1}
        \\
        \varphi_{1,2} & \varphi_{2,2} & \cdots & \varphi_{N,2}
        \\
        \vdots & \vdots & \ddots & \vdots
        \\
        \varphi_{1,L} & \varphi_{2,L} & \cdots & \varphi_{N,L}
    \end{pmatrix}.
\end{equation}
consisting of the molecular orbital coefficients of $N$ occupied molecular orbitals $\bm{\varphi}_i = \sum_{j=1}^L \varphi_{i,j} \, \bm{\chi}_j$ expanded in an orbital basis of $L$ one-electron functions $\{\bm{\chi}_j\}_{j=1}^L$. An $N$-electron Slater determinant (SD) state where $\bm{\varphi}_1, \bm{\varphi}_2, \ldots \bm{\varphi}_N$ are occupied can be written as
\begin{equation}\label{eqn:SD}
\begin{split}
        |\Psi^{\rm SD}\rangle &= \prod_{i=1}^N f^\dagger_{\bm{\varphi}_i} |{0}\rangle 
        \\
        &= \prod_{i=1}^N \left(\sum_{j=1}^L \varphi_{i,j} f^\dagger_{
        \bm{\chi}_j} \right) |{0}\rangle
        \\
        &= \sum_{1\leq j_1<j_2<\cdots<j_N\leq N}  \left(\sum_{\sigma \in S_{N}} {\rm sgn}(\sigma) \prod_{n=1}^N \varphi_{n,\,j_{\sigma(n)}} \right) f^\dagger_{\bm{\chi}_{j_1}} f^\dagger_{\bm{\chi}_{j_2}}\cdots f^\dagger_{\bm{\chi}_{j_N}} |{0}\rangle.
\end{split}
\end{equation}
where $f^\dagger_{\bm{\varphi}_i/\bm{\chi}_j}$ are fermionic creation operators which create electrons in orbitals $\bm{\varphi}_i/\bm{\chi}_j$ from the vacuum state $|{0}\rangle$, 
and $\sigma$'s are elements of the permutation group $S_N$. The last line of Eq.~\eqref{eqn:SD} reveals that the corresponding coefficient $\Psi^{\rm SD}(\bm{n})$ of an ONV state $|\bm{n}\rangle$ in the basis $\{\bm{\chi}_j\}_{j=1}^L$
\begin{equation}
    |\bm{n}\rangle = |n_1,n_2,\ldots,n_L\rangle \equiv \left(f^\dagger_{\bm{\chi}_1}\right)^{n_1}\left(f^\dagger_{\bm{\chi}_2}\right)^{n_2}\cdots\left(f^\dagger_{\bm{\chi}_L}\right)^{n_L} |{0}\rangle,
\end{equation}
is given by
\begin{equation}
    \Psi^{\rm SD}(\bm{n}) = \langle \bm{n} | \Psi^{\rm SD} \rangle = {\rm det}\, \bm{M}(\bm{n}),
\end{equation}
where $\bm{M}(\bm{n})$ is an $N\!\times\!N$ matrix consisting only of the $N$ rows of $\bm M$ with indices $i$'s such that $n_i = 1$. The coefficients $\Psi^{\rm SD}(\bm{n})$ are related by the same orbital coefficient matrix $\bm{M}$ such that $|\Psi^{\rm SD}\rangle$ is indeed a Slater determinant state. The key idea behind the NBF state $|\Psi^{\rm NBF}\rangle$ is that the coefficients of $|\bm{n}\rangle$ are instead computed from an \textit{ONV-dependent} set of occupied orbitals $(\bm{M}^{(\bm{n})})_{ij} = \varphi_{i,j}^{(\bm{n})}$, namely,
\begin{equation}
\begin{split}
    \Psi^{\rm NBF}(\bm{n}) = \langle \bm{n} | \Psi^{\rm NBF} \rangle &= {\rm det}\, \bm{M}^{(\bm{n})}(\bm{n}).
\end{split}
\end{equation}
As a result, the coefficients $\Psi^{\rm NBF}(\bm{n})$ can vary independently to introduce correlation, leading to a higher level of expressiveness. Additionally, by calculating the coefficients using backflowed determinants, NBF is able to impose a specific structure on the scalar output that gives a compact encoding of the target wavefunction. 

\begin{figure*}
    \centering
    \includegraphics[width=1\linewidth]{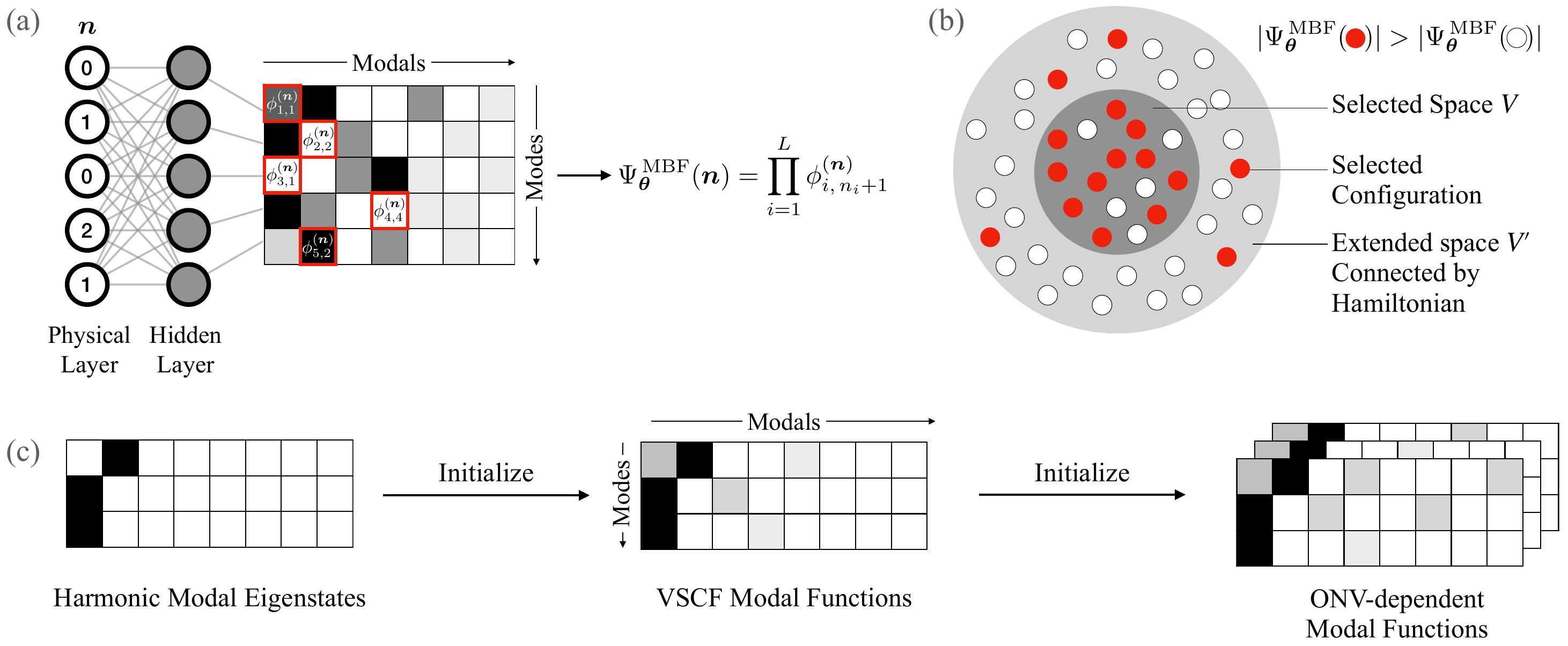}
    \caption{(a) Architecture of the MBF network: ONVs of a set of bosonic modes are fed through a shallow FNN and output to a set of modal functions, from which the wavefunction value is calculated based on the occupations of each mode. 
    (b) Selected-configuration scheme: For a given selected space $V$, a new selection of configurations can be proposed by selecting from the extended space $V'$ connected by the Hamiltonian, based on comparing the wavefunction amplitudes given by the current MBF wavefunction. 
    (c) Multi-step initialization scheme for boostraping the optimization of the MBF network: An eigenstate of the harmonic part of the Watson Hamiltonian is used to initialize a VSCF calculation. The resulting VSCF solution is used as the fixed part of the ONV-dependent modal functions, while the actual ONV-dependent deviations are learned by the MBF network.}
    \label{fig:mbf_arch}
\end{figure*}

However, the generalization of NBF to the case of bosonic modes is not straightforward. The direct analogue of the backflow determinant for bosons would be the backflow \textit{permanent}, because bosonic wavefunctions must be fully symmetric. Although the matrix determinant can be computed with polynomial cost, no such efficient algorithm is available to exactly compute the matrix permanent that is needed for the symmetry of bosons.\cite{valiant1979complexity} Moreover, in many bosonic problems of interest, the number of bosons is not a good quantum number. This means that the number of bosons involved in each configuration and, therefore, the size of the matrix of which the permanent is to be computed is not fixed.

To avoid the issues above, we propose an alternative formulation of neural backflow ansatz using modals.\cite{christiansen2004second} Modals are elements of the Fock space of a single mode $i$, the canonical example being the eigenstates $|n_i\rangle = (b^\dagger_i)^{n_i}|0\rangle$ of the particle number operator $\hat{N}_i = b^\dagger_i b^{\phantom{\dagger}}_i$, where $n_i = 0, 1, \ldots, N_{\rm max}$. In general, a modal wavefunction $|\bm{\phi}_i\rangle$ is an arbitrary superposition
\begin{equation}
    |\bm{\phi}_i\rangle = \sum_{n_i=0}^{N_{\rm max}} \phi_{i,n_i+1} |n_i\rangle,
\end{equation}
where $\bm{\phi_i} = (\phi_{i,j})_{j=1}^{N_{\rm modal}}$ (recall that $N_{\rm modal}=N_{\rm max}+1$) is a vector of modal coefficients. An $L$-mode state without any mode-mode correlation, such as the vibrational self-consistent field (VSCF) \cite{bowman1986self,gerber1988self} solution, is a modal product (MP) state,
\begin{equation}
\begin{split}
    |\Psi^{\rm MP}\rangle &= \bigotimes_{i=1}^L |\bm{\phi}_i\rangle = \bigotimes \left(\sum_{n_i=0}^{N_{\rm max}} \phi_{i,n_i+1} |n_i\rangle \right)  = \sum_{\bm{n}} \left( \prod_{i=1}^L \phi_{i,n_i+1} \right) |\bm{n}\rangle,
\end{split}
\end{equation}
where
\begin{equation}
    |\bm{n}\rangle = |n_1,n_2,\ldots,n_L\rangle \equiv (b^\dagger_{1})^{n_1}(b^\dagger_{2})^{n_2}\cdots(b^\dagger_{L})^{n_L} |{0}\rangle.
\end{equation}
The coefficient $\Psi^{\rm MP}(\bm{n})$ of a bosonic ONV state $|\bm{n}\rangle$ is then given by
\begin{equation}
    \Psi^{\rm MP}(\bm{n}) = \langle \bm{n} | \Psi^{\rm MP}\rangle  = \prod_{i=1}^L \phi_{i,n_i+1},
\end{equation}
without the need for computing the permanent of a matrix.
Analogously, to incorporate correlation into the wavefunction, we introduce the modal backflow (MBF) state $|\Psi^{\rm MBF}\rangle$ to be such that the modal functions $\bm{\phi}_i$'s are ONV-dependent. Accordingly, the coefficients of the ONV states become
\begin{equation}
    \Psi^{\rm MBF}(\bm{n}) = \langle \bm{n} | \Psi^{\rm MBF}\rangle =  \prod_{i=1}^L \phi^{(\bm{n})}_{i,n_i+1}.
\end{equation}
The $L$ ONV-dependent modal functions $\bm{\phi}_i^{(\bm{n})}$ form an $L\!\times\!N_{\rm modal}$ matrix which we will learn with a fully connected neural network. The MBF state naturally covers all bosonic particle number sectors, and correlates both within and among different sectors freely, without incurring the exponential computational cost of evaluating the bosonic permanent. Moreover, the modal representation allows us to work with a matrix of fixed dimensions consisting of $L$ modal functions. We summarize the MBF network architecture in Figure \ref{fig:mbf_arch} (a).

The parameters $\bm{\theta} = (\bm{W},\bm{b})$ of the MBF wavefunctions are the weights $\bm{W}$ and biases $\bm{b}$ that propagate the information through the network according to
\begin{equation}
    \begin{split}
        \bm{h} = \tanh (\bm{W}^{(0)} \bm{n} + \bm{b}^{(0)}),
        \\
        \bm{o} = \tanh (\bm{W}^{(1)} \bm{h} + \bm{b}^{(1)}),
    \end{split}
\end{equation}
where $\bm{h}$ and $\bm{o}$ are the hidden and output layers, respectively. The vector $\bm{o}$ is flattened from the output matrix consisting of $L$ modal functions of length $N_{\rm modal}$ depicted in Figure \ref{fig:mbf_arch} (a). The number of hidden neurons $N_{\rm hidden}$ can be regulated using the hidden neuron density $\alpha = N_{\rm hidden}/L$, which also acts as a tuning parameter for the complexity of the network. In practice we break down the ONV-dependent modal functions into a fixed part (which corresponds to a mode product state), and the variable part that is actually ONV-dependent
\begin{equation} \label{eqn:bf_modals}
    \bm{\phi}_i^{(\bm{n})} = \bm{\phi}_i^{(0)} + \bm{\delta}_i^{(\bm{n})}.
\end{equation}
During training, $\bm{\delta}_i^{(\bm{n})}$ is reshaped from the vector output $\bm{o}$ of the network. In a later section, we will describe the way the fixed modal functions $\bm{\phi}_i^{(0)}$ are chosen.

\subsection{Optimization}

In this section, we discuss several aspects of the optimization of the MBF network that are crucial to achieving the target accuracy. The implementation of MBF is carried out using the NQS package Netket.\cite{netket2:2019,netket3:2022}

\textbf{Markov Chain Monte Carlo.} The most common method of extracting observables from a neural quantum state $\Psi$ is by Monte Carlo sampling.\cite{mcmillan1965ground,foulkes2001quantum} The expectation of an observable is mathematically equivalent to a weighted sum of the so-called local energy $O_{\rm loc}$,
\begin{equation}
\begin{split}
    \langle \hat{O} \rangle_{\Psi} &= \frac{\langle \Psi|\hat{O}|\Psi\rangle}{\langle \Psi | \Psi\rangle}=\sum_{\bm{n}} P_\Psi(\bm{n}) O_{\rm loc}(\bm{n}),
\end{split}
\end{equation}
with
\begin{equation}
\begin{split}
    P_\Psi(\bm{n})&\equiv \frac{|\Psi(\bm{n})|^2}{\sum_{\bm{n}'} |\Psi(\bm{n}')|^2 },
    \end{split}
\end{equation}
and
\begin{equation}
\begin{split}
    O_{\rm loc}(\bm{n}) &\equiv \sum_{\bm{n}'} \langle \bm{n} | \hat{O} |\bm{n}'\rangle \frac{\Psi(\bm{n}')}{\Psi(\bm{n})},
\end{split}
\end{equation}
where $P_\Psi(\bm{n})$ is the probability distribution given by the quantum state $\Psi$. In Monte Carlo sampling, the exact weighted sum over all configurations $\bm{n}$ is approximated by a Markov chain $\mathcal{C}$ of configuration samples drawn according to probability distribution $P_\Psi(\bm{n})$. In other words,
\begin{equation}
\begin{split}
    \sum_{\bm{n}} P_\Psi(\bm{n}) O_{\rm loc}(\bm{n}) \approx \sum_{\bm{n} \in \mathcal{C}} O_{\rm loc} (\bm{n}).
\end{split}
\end{equation}
Although this is a simple and powerful method, its stochastic nature does not align well with the requirement of high precision in spectroscopic calculations. Moreover, in most scenarios anharmonic eigenstates are dominated by a single configuration, resulting in a sharp peak in the probability distribution which heavily skews the sampling. One needs an exceedingly large number of samples in order to reach enough configurations beside the dominating one.

\textbf{Stochastically Selected Configuration.} To avoid the above issues, we use a selected-configuration scheme introduced in Ref.~\cite{li2023nonstochastic}, which is illustrated in Figure \ref{fig:mbf_arch} (b). Instead of replacing the exact weighted sum over all configurations with Monte Carlo sampling, we restrict the weighted sum to only a small number of distinct configurations with the highest weights according to the full wavefunction, collected in the set $V$.
Mathematically, this is equivalent to an asymmetric evaluation of the expectation with the truncated state $|\Psi_{V}\rangle = \sum_{\bm{n} \in V} \Psi(\bm{n})|\bm{n}\rangle$ and the exact state
\begin{equation}
\begin{split}
    \sum_{\bm{n}} P_\Psi(\bm{n}) O_{\rm loc}(\bm{n}) &\approx \sum_{\bm{n} \in V} P_{\Psi_V}(\bm{n}) O_{\rm loc} (\bm{n}) 
    \\
    &= \frac{\langle \Psi_V|\hat{O}|\Psi\rangle}{\langle \Psi_V|\Psi_V\rangle} 
    \\
    &\equiv \langle \hat{O} \rangle_{V}.
\end{split}
\end{equation}
Note that the evaluation $\langle \cdot \rangle_{V}$ is no longer strictly variational. However, we will find that the variational condition is well maintained as long as all configurations with significant weights are selected in $V$. 

Optimizing an NQS $\Psi_{\bm{\theta}}$ with network parameters $\bm \theta$ using the selected-configuration scheme consists of the following steps:
\begin{enumerate}
    \item Initialize $\Psi_{\bm{\theta}^{(0)}}$ and $V^{(0)}$ as $\{\bm{n}_0\}$ where $\bm{n}_0 = (0,0,\ldots,0)$, or another seed state of choice (depending on the level of excitation).
    \item In step $t$, compute the set $V' = \hat{H}V^{(t)}$ consisting of configurations connected by $\hat{H}$ to $V^{(t)}$. 
    \item Select $N_s$ configurations from $V^{(t)} \cup V'$ with the highest amplitudes $|\Psi_{\bm{\theta}}(\bm{n})|$ to form the new set of selected states $V^{(t+1)}$.
    \item Evaluate the energy gradients $\partial E / \partial \theta_m$ and update the network parameters $\bm{\theta}^{(t)}\leftarrow \bm{\theta}^{(t+1)}$.
    \item Repeat from step 2 until the maximum number of iterations is reached.
\end{enumerate}
The above scheme was first implemented to solve the electronic Hamiltonian.\cite{li2023nonstochastic}
In our case, the vibrational Hamiltonian contains long strings of creation and annihilation operators that go beyond the two-body terms. The implication is that the connected subspace $V'$ in the vibrational case can be prohibitively large, making the step of identifying the $N_s$ states with the highest contributing weights an expensive task. To avoid this, we restrict the full extended space $V'$ to a smaller subset consisting of only $K\!\times\! N_s$ randomly sampled connected states, where $K$ can be tuned as a parameter of the degree of exploration. This randomized restriction not only eases the computational effort, but also introduces some level of stochasticity. 

Within the selected-configuration scheme, the gradients of the expectation value $\langle \hat{O} \rangle_V$ with respect to the network parameters $\theta_m$'s are defined as
\begin{equation}
\begin{split}
    F_m = \frac{\partial E}{\partial \theta_m} = \sum_{\bm{n} \in V} P_{\Psi}(\bm{n}) D_m(\bm{n})^\ast(O_{\rm loc}(\bm{n}) - \langle \hat{O}\rangle_V),
\end{split}
\end{equation}
where
\begin{equation}
    D_m(\bm{n}) = \frac{1}{\Psi^{\rm MBF}_{\bm{\theta}}(\bm{n})} \frac{\partial \Psi^{\rm MBF}_{\bm{\theta}}(\bm{n})}{\partial \theta_m},
\end{equation}
is the derivative of the logarithm of the wavefunction.
At each iteration, we update the network parameters $\bm \theta$ using the continuous resilient (CoRe) optimizer \cite{eckhoff2023,eckhoff2024core}
\begin{equation}
    \bm{\theta}^{(t+1)} = \bm{\theta}^{(t)} - \eta\, \bm{G}^{(t)},
\end{equation}
where $\eta$ is the learning rate and $\bm{G}^{(t)}$ is the scaled gradient
\begin{equation}
\begin{split}
    {G}^{(t)}_m &= \frac{{g}_m^{(t)}}{1-(\beta_1^{(t)})^t} \left( \sqrt{\frac{{h}_m^{(t)}}{1-\beta_2^{\,t}}} + \epsilon \right)^{-1}
    \\
    g^{(t)}_m &= \beta_1^{(t)} g^{(t-1)}_m + (1-\beta_1^{(t)}) F_m^{(t-1)},
    \\
    h^{(t)}_m &= \beta_2\,\, h^{(t-1)}_m + (1-\beta_2)\left(F_m^{(t-1)}\right)^2,
    \\
    \beta_1^{(t)} &= \beta_1^b + (\beta_1^a - \beta_1^b) \exp\left[-\left(\frac{t-1}{\beta_1^c}\right)^2\right].
\end{split}
\end{equation}
The default hyperparameters used in this work are $\eta=0.05$, $\beta_1^a=0.9$, $\beta_1^b = 0.5$, $\beta_1^c=100$, $\beta_2=0.99$.

\textbf{VSCF pretraining.} Pretraining the network parameters is a proven technique for accelerating and stabilizing the optimization of the network.\cite{pfau2020ab,liu2024backflow} Here, we can pretrain the backflow-free part $\bm{\phi_i}^{(0)}$ of the ONV-dependent modal functions $\bm{\phi_i}^{(\bm{n})}$ in Eq.~\eqref{eqn:bf_modals} to match the solution of a vibrational self-consistent field (VSCF) theory.\cite{bowman1986self,gerber1988self} The VSCF theory assumes a modal product (MP) state, where each mode $i$ is described by a single modal function $\bm{\phi}_i$,
\begin{equation}
    |\Psi^{\rm MP}\rangle = \bigotimes_{i=1}^L |\bm{\phi}_i\rangle_i.
\end{equation}
The mean-field solution to the Watson Hamiltonian is then obtained by the minimization
\begin{equation}
\begin{split}
    E^{\rm VSCF} &= \min_{|\Psi^{\rm MP}\rangle} \frac{\langle \Psi^{\rm MP} | \hat{H}_{\rm vib} | \Psi^{\rm MP}\rangle }{\langle \Psi^{\rm MP} | \Psi^{\rm MP}\rangle }
    \\
    &=  \frac{\langle \Psi^{\rm VSCF} | \hat{H}_{\rm vib} | \Psi^{\rm VSCF} \rangle}{\langle \Psi^{\rm VSCF} | \Psi^{\rm VSCF} \rangle}.
\end{split}
\end{equation}
This minimization is equivalent to solving for the optimal set of $\bm{\phi}^{(0)}_i$, which is, in turn, equivalent to optimizing the MBF network with all weights and biases set to zero (meaning $\bm{\delta}^{(\bm{n})}_i = \bm{0}$). The VSCF modal functions already capture a considerable amount of anharmonic correction to the energy, allowing us to focus on the remaining part of the energy minimization by optimizing only the ONV-dependent modal corrections $\bm{\delta}^{(\bm{n})}_i$. The optimization for $\bm{\phi_i}^{(0)}$ is itself initialized with the eigenstates of the harmonic part of the vibrational Hamiltonian, which are ONV states (e.g., the vacuum for the ground state). The assumption behind this initialization scheme is that the harmonic solutions are in the vicinity of the VSCF solutions, which are themselves in the vicinity of the anharmonic solutions. We illustrate this nested pretraining procedure in Figure \ref{fig:mbf_arch} (c). The VSCF modal functions are optimized using the same selected-configuration procedure, and the final set of selected-configurations are fed to the MBF network together with the VSCF modal functions. We will see that the VSCF pretraining significantly speeds up the optimization of the MBF wavefunction, and is crucial to avoiding getting stuck in higher lying states when targeting excited states.

\textbf{Targeting excited states.} There are two main approaches to calculate excited states: state-average/ensemble methods \cite{entwistle2023electronic,pfau2024accurate} and state-specific calculations using an orthogonality penalty.\cite{choo2018symmetries,pathak2021excited,wei2024finding} In this work we choose the second route for its simple implementation. Specifically. we use the shifted Hamiltonian and the corresponding implementation of the excited state solver using the NQS package NetKet \cite{netket2:2019,netket3:2022} given in Ref.~\cite{wei2024finding}. The penalty-modified Hamiltonian for the $n$-th excited state reads
\begin{equation}
    \hat{H}_{\rm vib}^{(n)}(z) = \hat{H}_{\rm vib} + z \sum_{j=1}^{n-1} \frac{|\Psi_j\rangle \langle \Psi_j|}{\langle \Psi_j |\Psi_j\rangle}- {\rm ZPE}, \quad n \geq 1 ,
\end{equation}
where the zero-point energy (ZPE) is subtracted for convenience.
Mathematically, the shift constant $z$ can be chosen to be any value higher than the target vibrational transitions. For example, setting $z = 1000\, \rm cm^{-1}$ allows us to target all vibrational transitions from the ZPE below 1000 $\rm cm^{-1}$. In practice, however, a $z$ value that is too high can skew the optimization to prioritize maintaining orthogonality rather than minimizing the excited state energy. Therefore, $z$ should ideally be level-dependent and only slightly above the excited state energy, which naturally requires some \textit{a priori} knowledge of the target energy level. Since the $n$-th exact eigenstate of $\hat{H}_{\rm vib}$ is also the variational ground state of $\hat{H}_{\rm vib}^{(n)}$ regardless of the value of $z$, we can estimate $z$ to be the energy of the VSCF solution of $\hat{H}_{\rm vib}^{(n)}$
\begin{equation}
\begin{split}
    z_n &= \frac{\langle \Psi^{\rm VSCF}_n | \hat{H}_{\rm vib} |\Psi^{\rm VSCF}_n \rangle}{\langle \Psi^{\rm VSCF}_n | \Psi^{\rm VSCF}_n \rangle} + z' \sum_{j=1}^{n-1} \frac{|\langle \Psi_j | \Psi_{n}^{\rm VSCF}\rangle|^2}{\langle \Psi_j |\Psi_j\rangle\langle \Psi^{\rm VSCF}_n |\Psi^{\rm VSCF}_n \rangle} - {\rm ZPE}
    \\
    &\geq\frac{\langle \Psi_n^{\rm exact}|\hat{H}_{\rm vib} | \Psi^{\rm exact}_n\rangle}{\langle \Psi_n^{\rm exact}| \Psi^{\rm exact}_n\rangle} - {\rm ZPE} = E_n - {\rm ZPE},
\end{split}
\end{equation}
where $z'$ is an initial estimate of the shift constant.
Typically, the overlap $|\langle\Psi_j|\Psi_n^{\rm VSCF}\rangle|$ is already small due to the fact that they are dominated by different harmonic configurations, and $z_n$ is considerably improved compared to $z'$.

\section{Numerical Experiment}\label{sec:num}
In this section, we investigate different degrees of anharmonicity to be described by the MBF network. In order to allow for a systematic investigation, we decided on creating an artifical Hamiltonian for a 4-mode system with randomly sampled third- and fourth-order partial derivatives $\kappa^{(\nu)}_{i_1,i_2,\ldots,i_\nu}$ ($\nu = 3,4$) of the PES that emulate weak, moderate, and strong anharmonicity. The harmonic frequencies are random numbers uniformly sampled from the interval [1500, 3000] $\rm cm^{-1}$ and the partial derivatives of the PES are sampled according to normal distributions
\begin{equation}
    |\kappa^{(\nu)}_{i_1,i_2,\ldots,i_\nu}| = \mathcal{N}(\lambda_\nu, \lambda_\nu/5), \quad \nu = 3,4,
\end{equation}
where the mean $\lambda$ and standard deviation $\lambda_\nu/5$ for the normal distribution $\mathcal{N}$ are controlled by the parameter (in Hartree atomic unit, a.u.)
\begin{equation}
    \lambda_\nu = \begin{cases}
        (50{\rm \: cm^{-1}})_{\rm a.u.} \times \bar{w}^{\,\nu/2}, \quad &\rm{weak}
        \\
        (150{\rm \: cm^{-1}})_{\rm a.u.} \times \bar{w}^{\,\nu/2}, \quad &\rm{moderate}
        \\
        (500{\rm \: cm^{-1}})_{\rm a.u.} \times \bar{w}^{\,\nu/2}, \quad &\rm{strong}
    \end{cases}
\end{equation}
for weak, moderate, and strong anharmonicity. Here, $\bar{w}$ was taken to be the midpoint of the sampling interval (2250 cm$^{-1})_{\rm a.u.}$. Furthermore, to mimic a typical PES, we took the absolute values of the fourth-order constants and set an off-diagonal decay factor $\eta=1,0.1,0.01$ for fully- (all indices identical), semi- (only some indices identical), and off-diagonal (no indices identical) tensor elements, respectively. Throughout the numerical experiment, only cubic and quartic force constants were used for the Watson Hamiltonian, and no Coriolis terms were included.

\begin{figure}
    \centering
    \includegraphics[width=0.7\linewidth]{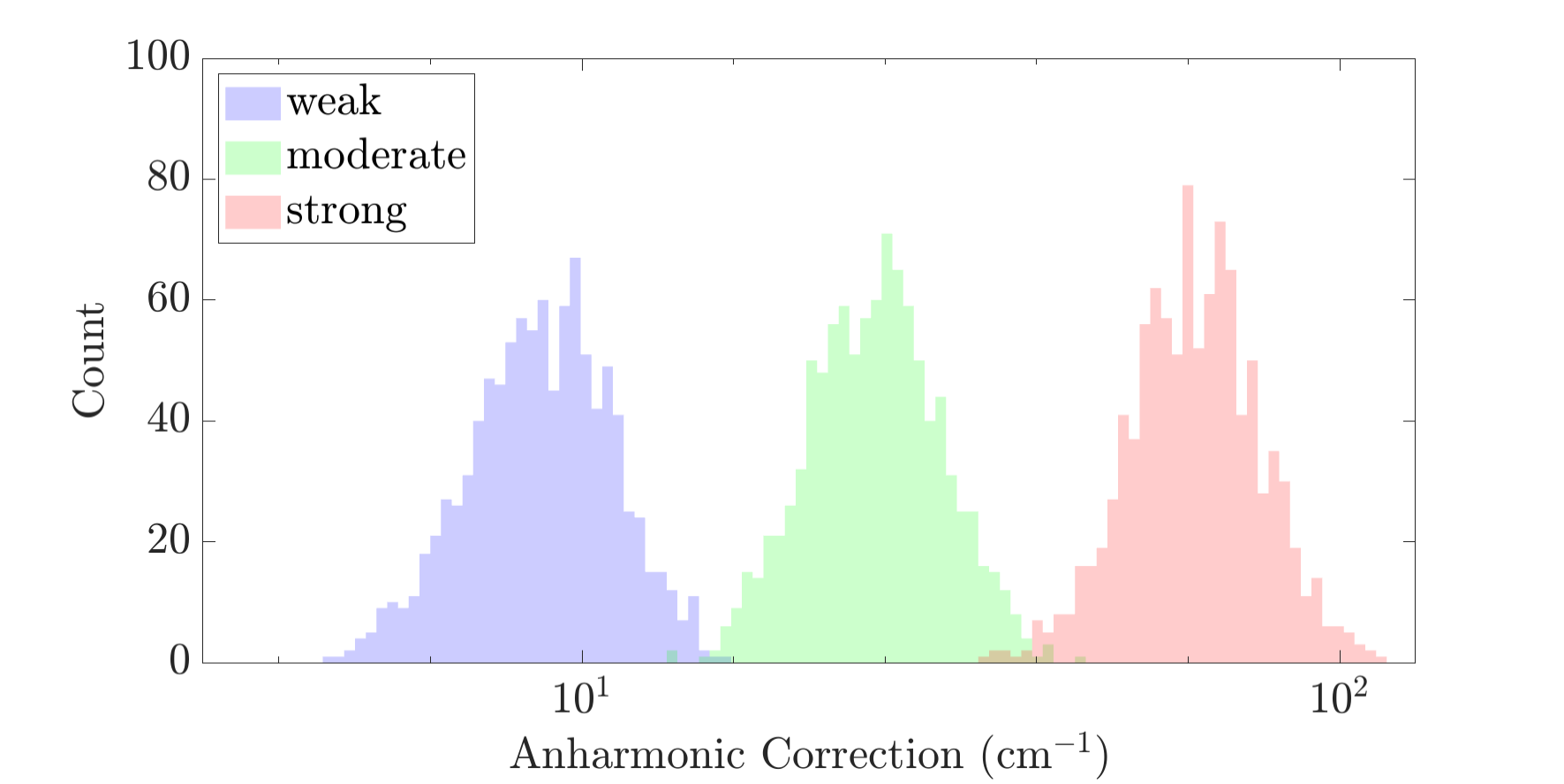}
    \caption{Distribution of the anharmonic correction of sampled anharmonic 4-mode Hamiltonians with $N_{\rm max}=9$. In each distribution, $10^3$ Hamiltonians were sampled, and both $\kappa^{(3)}_{ijk}$ and $\kappa^{(4)}_{ijkl}$ were set to the weak, moderate, and strong regime of anharmonicity.}
    \label{fig:vibham_dist}
\end{figure}

\begin{figure}
    \centering
    \includegraphics[width=0.7\linewidth]{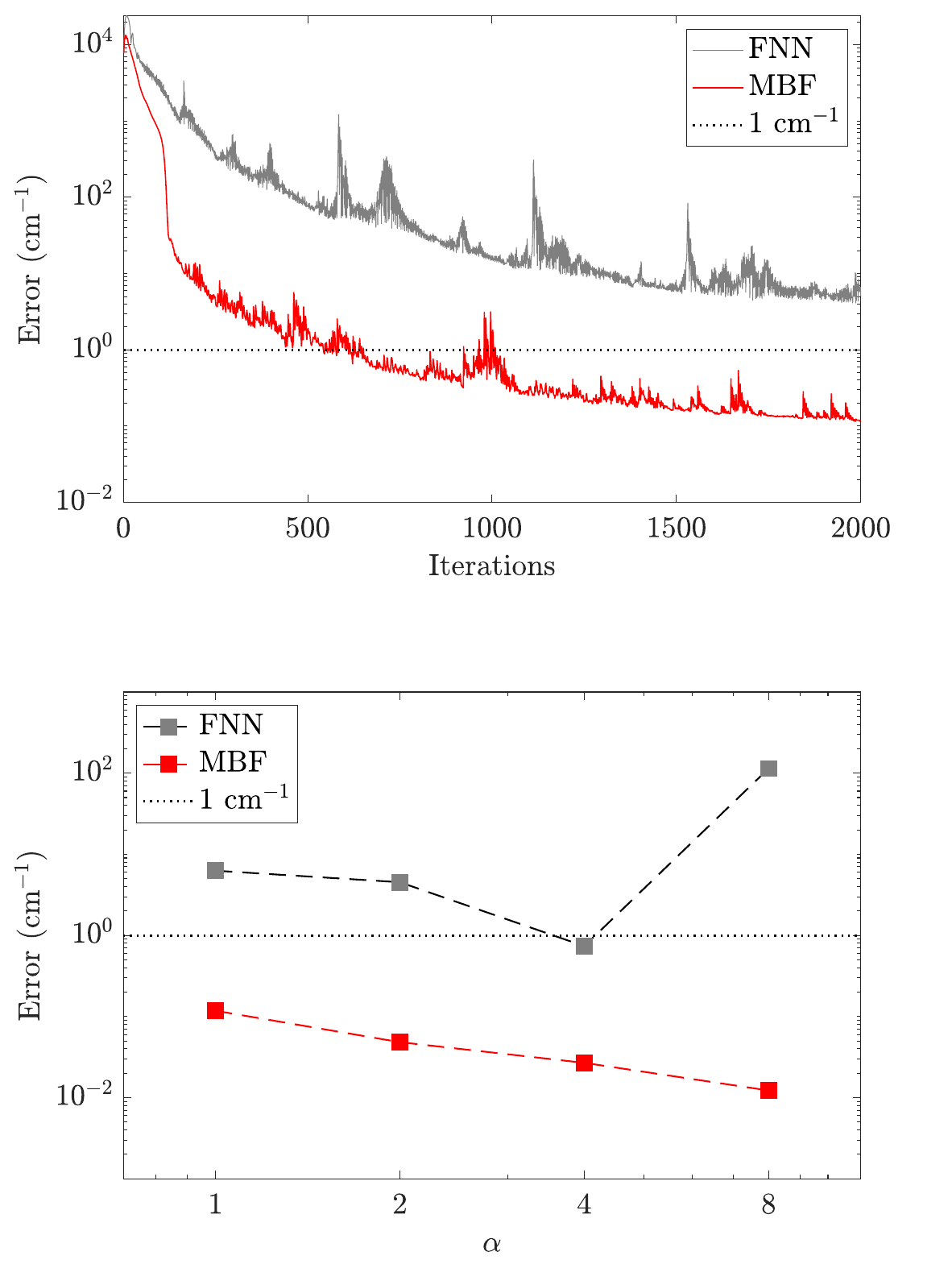}
    \caption{(Top) Comparison of the optimization between FNN and MBF for targeting the ground state of a randomly generated 4-mode Watson Hamiltonian with the moderate anharmonicity setting. For both networks we set $\alpha=1$. (Bottom) Comparison of the final error between FNN and MBF after 2000 iterations, for $\alpha = 1,2,4,$ and 8. $N_s=128$ and $N_{\rm max}=6$ for all calculations. }
    \label{fig:fnn_compare}
\end{figure}

\begin{figure*}
    \centering
    \includegraphics[width=0.9\linewidth]{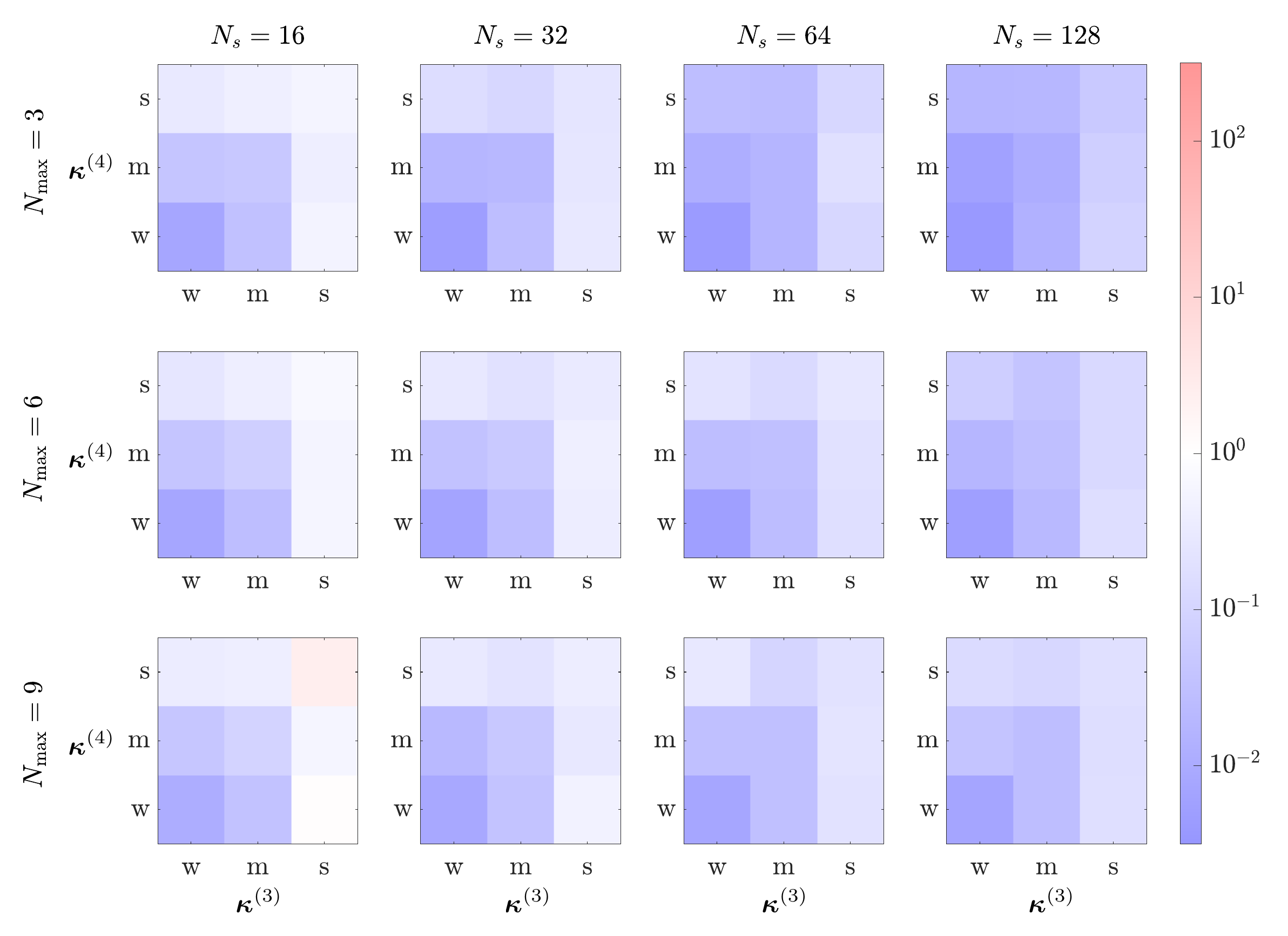}
    \caption{Error (cm$^{-1}$) of the ground state optimization with MBF wavefunctions. Each 3$\times$3 block correspond to a combination of $N_{\rm max}$ and selected space dimension $N_s$. Within each block, different combinations of anharmonic strength (w: weak, m: moderate, s: strong) of third- ($\bm{\kappa}^{(3)}$) and fourth-order partial derivatives ($\bm{\kappa}^{(4)}$) of the PES are used to generate 100 4-mode Watson Hamiltonians, over which the final errors of the MBF energy are averaged and color-coded on a log-scale, with red, white, and blue to represent above, at, and below spectroscopic accuracy of 1 cm${-1}$, respectively. All calculations used the same hyperparameters for the optimization and ran for 1000 iterations. A VSCF pretraining was used before each optimization.}
    \label{fig:error_stat}
\end{figure*}

We quantified the anharmonicity of the sampled Hamiltonians by the anharmonic correction in their ZPE, defined as the difference between the ground state energy of $\hat{H}_{\rm harm} = \sum_{i=1}^L w_i (\hat{n}_i + \frac{1}{2})$ and that of $\hat{H}_{\rm vib}$. In Figure \ref{fig:vibham_dist} we plotted the distribution anharmonic correction of $10^3$ sampled Hamiltonian, with $\kappa^{(3)}_{ijk}$ and $\kappa^{(4)}_{ijkl}$ both set to represent weak, moderate, and strong anharmonicity. 
For each Hamiltonian, the ground state was calculated with exact diagonalization and $N_{\rm max}$ was set to 9. We found that the three settings indeed cover different ranges of the anharmonic correction with very little overlap. The setting for strong anharmonicity reaches the order of a 100 cm$^{-1}$ anharmonic correction, which already surpasses molecules considered to be strongly anharmonic. 

First, we investigate the effect of learning ONV-dependent modal functions instead of directly learning the wavefunction values through a shallow FNN. For this purpose, we chose the moderate setting for both the third- and fourth-order reduced force constants, and set to $N_{\rm max} = 6$. To isolate the effect of the network architecture for our comparison, we fixed $N_s = 128$ and did not apply any pretraining. 
The size of the extended subspace $V'$ was set at $5N_s$. Furthermore, we imposed an exponential decay in the learning rate $\eta=\exp(-rt/T)\eta_0$, where $t$ and $T$ are the current and total step counts, respectively, and the decay rate $r$ was set to 0.1.

In the top panel of Figure \ref{fig:fnn_compare}, we show a comparison between the accuracy reached by the FNN and MBF networks after 2000 iteration steps, both with a hidden neuron density $\alpha = 1$. 
The energy of the MBF wavefunction improved significantly faster than that of FNN. 
At the end of the optimization, the error of the MBF energy reached around 0.1 cm$^{-1}$, which is two orders of magnitude smaller than that of the FNN energy. In the bottom panel of Figure \ref{fig:fnn_compare}, we compared the final errors of the energy reached by both networks with different values of hidden neuron density $\alpha$. As $\alpha$ increases, the error of the MBF energy decreases roughly according to a power law, indicated by the approximately linear curve in the logarithmic scale. We first found a similar improvement in FNN. At $\alpha=4$, the FNN was able to reach spectroscopic accuracy of 1 cm$^{-1}$ and lower. However, increasing $\alpha$ further made the optimization unstable and did not produce an improved error. Together, both plots in Figure \ref{fig:fnn_compare} demonstrate the clear advantage of the MBF network over FNN in terms of expressiveness and systematic improvability.

Next, we analyzed the reliability of MBF for different combinations of anharmonic strength of the third- ($\bm{\kappa}^{(3)}$) and fourth-order partial derivatives ($\bm{\kappa}^{(4)}$) and for different $N_{\rm max}$. For each combination, we used an MBF network with hidden neuron density $\alpha = 1$, and the sizes of the selected subspaces were $N_s = 16, 32, 64,$ and $128$. In Figure \ref{fig:error_stat}, we showed the errors of the MBF ground state energy for different anharmonic strengths, grouped by $N_{\rm max}$ and $N_s$. The errors were measured against exact diagonalization results and color coded on a logarithmic scale in red, white, and blue to represent above, on, and below the spectroscopic accuracy of 1 cm$^{-1}$, respectively. For each combination of $(\bm{\kappa}^{(3)},\bm{\kappa}^{(4)},N_{\rm max},N_s)$, 100 Hamiltonians were sampled and solved, and the final errors were averaged. A VSCF pretraining was used for all calculations. Almost all errors were below 1 cm$^{-1}$ (blue), except for the top right corner of the grid of $N_{\rm max} = 9$ and $N_s = 16$, which corresponds to solving the largest and most anharmonic system with the smallest number of selected-configurations.  
Moreover, we observed the following trends: (1) increasing $N_s$ systematically reduces the final error; (2) the stronger the anharmonicity, the more configurations should be taken into the selected space to increase accuracy; (3) a larger $N_{\rm max}$ typically also calls for a larger selected subspace to reach spectroscopic accuracy of 1 cm$^-1$. 

In this section, using randomly generated Watson Hamiltonians, we systematically tested the performance of the MBF network in terms of its advantages over FNNs, its range of applicability, and the effect of the choices of parameters such as the hidden neuron density $\alpha$ and size of selected space $N_s$. The insights gained provide valuable guidance as we now apply the MBF network to solve for the ground and excited states of \textit{ab initio} anharmonic vibrational Hamiltonians of molecules.

\section{Ab Initio Anharmonic Vibrational Hamiltonians}\label{sec:ab}

We applied the MBF network to target the ZPE and low-lying vibrational transitions of three molecules, $\rm ClO_2$, $\rm H_2CO$, and $\rm CH_3CN$, increasing both the number of modes and the degree of anharmonicity. For $\rm ClO_2$ and $\rm H_2CO$, we used the sextic force field and Corioli constants from the library PyPES \cite{pypes}. For the $\rm CH_3CN$ molecule, although only quartic force fields are available (e.g., see  Refs.~\cite{begue2005calculations,avila2011using}), its strong anharmonic character makes it a suitable subject for benchmarking tensor network methods.\cite{baiardi2017vibrational,larsson2025benchmarking}. 
Here, we used the PES reported in Ref.~\cite{avila2011using}, which was directly taken from the Supporting Information of Ref.~\cite{larsson2025benchmarking}. The reference energies were computed using the program QCMaquis \cite{szenes2025qcmaquis} with the vibrational density matrix renormalization group (vDMRG) calculations. 
We chose MPSs of maximum bond dimensions 100, which is sufficiently large for the largest system $\rm CH_3CN$ according to Ref.~\cite{baiardi2017vibrational}. For the MBF calculations, the size of the extended subspace $V'$ was set at $2N_s$ and the exponential decay rate $r$ of the learning rate $\eta$ is set to 0.5.

\begin{figure}
    \centering
    \includegraphics[width=0.7\linewidth]{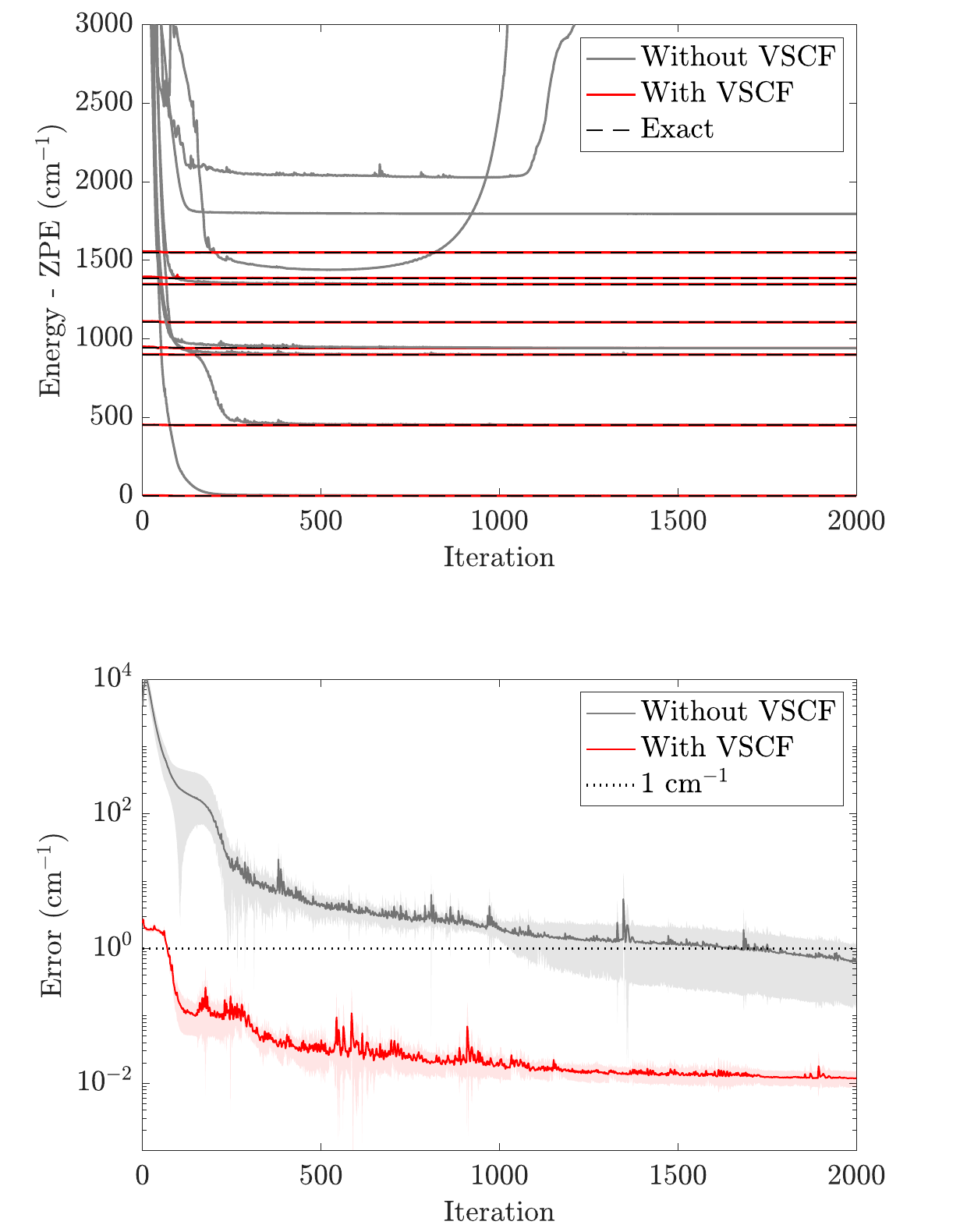}
    \caption{(Top) Optimization of the lowest eight vibrational levels (shifted by the ZPE) of $\rm ClO_2$ using an MBF with $\alpha = 2$ and $N_s = 64$. Energy levels obtained by vDMRG are plotted as reference. $N_{\rm max}=9$. (Bottom) Comparison of the average error of the three lowest eigenstates between optimizations with and without VSCF pretraining. The shaded area marks the range of the standard deviation of the three states.}
    \label{fig:clo2}
\end{figure}

The triatomic $\rm ClO_2$ is with its three normal modes weakly anharmonic. First, we demonstrate the ability of the VSCF pretraining step in accelerating and stabilizing excited state optimizations. In the top panel of Figure \ref{fig:clo2}, we show the optimization of an MBF ($\alpha = 2$ and $N_s = 64$) targeting the lowest eight eigenstates, both with and without VSCF pretraining. 
When the optimization starts from a random initial guess without VSCF pretraining, the optimization for the ground and first two excited states still converged, albeit slowly. However, from the third excited state onwards, the optimization tends to get trapped in other local minima, eventually arriving at the wrong eigenstates. The third excited state, for example, was only obtained in the fifth calculation.
Some optimizations even became unstable and failed to reach any of the eight low-lying states. 

When VSCF pretraining was employed, the optimization started already around the exact eigenenergies, and its improvement is hardly visible on the scale chosen. A superior set of initial modal functions given by the VSCF pretraining
also ensured that we obtain each eigenstate in the correct order, which is a crucial feature for identifying all transition energies below the target threshold. For the lowest three eigenstates, for which optimizations both with and without VSCF pretraining converged, we compared the decaying behavior of the error averaged over three states in the bottom panel of Figure \ref{fig:clo2}. We found that optimizing with VSCF pretraining allowed the error to reach 1 cm$^{-1}$ accuracy much faster than in the case without pretraining, and eventually this reduced the error by about two orders of magnitude. The accuracy was well maintained from ground to excited states when VSCF pretraining was in place, indicated by the small standard deviation (pink shaded area) compared to the case without pretraining (gray shaded area).

\begin{figure}
    \centering
    \includegraphics[width=0.7\linewidth]{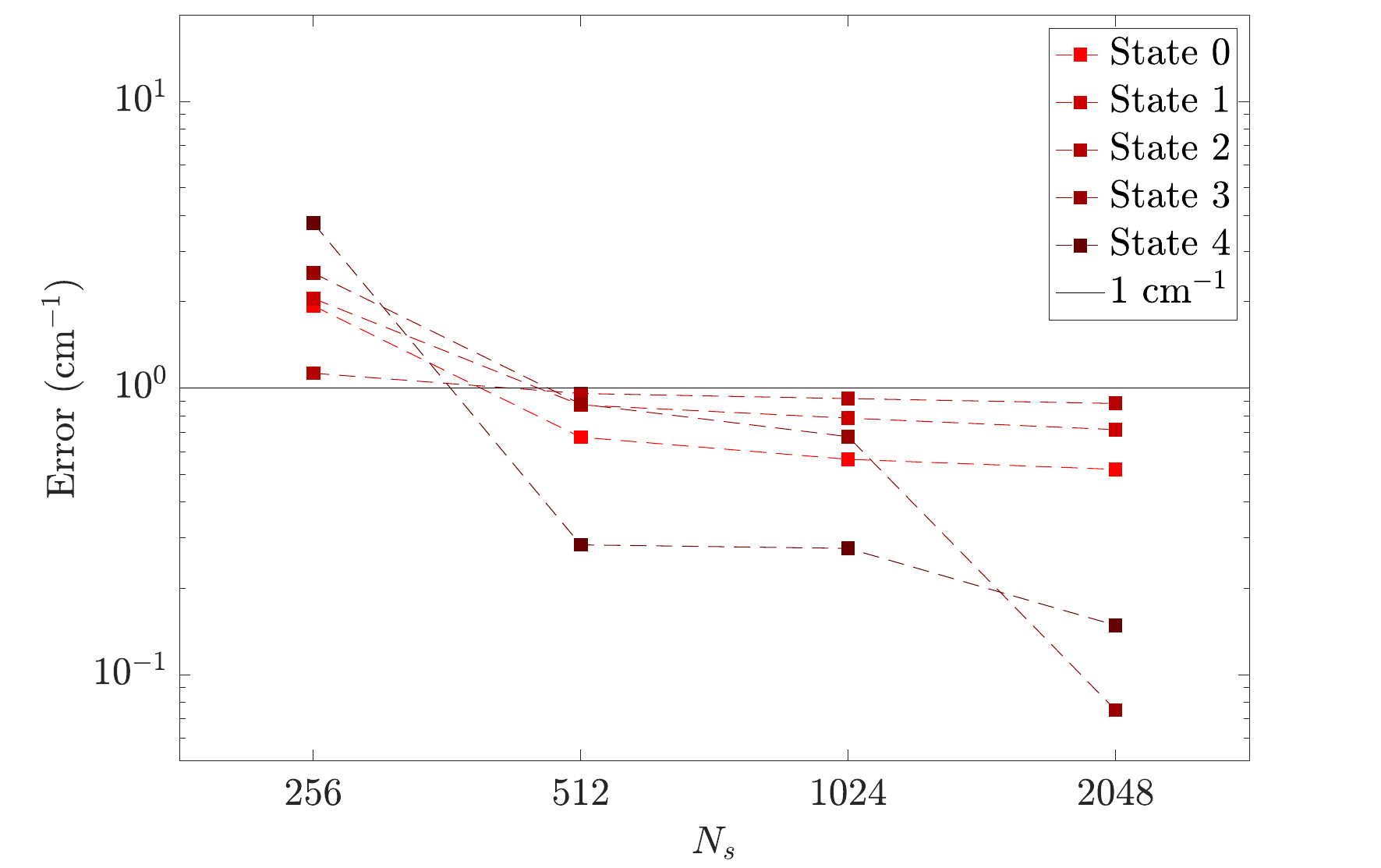}
    \caption{Error (cm$^{-1}$) of the eigenstate energies of $\rm H_2CO$ calculated with MBF and compared to vDMRG results as $N_s$ increases ($N_{\rm max}=6$).}
    \label{fig:h2co}
\end{figure}

For the larger molecules $\rm H_2CO$ (6 modes) and $\rm CH_3CN$ (12 modes) with $N_{\rm max}=6$ and increased anharmonicity, indicated by the anharmonic corrections of about 77 cm$^{-1}$ for $\rm H_2CO$ and 67 cm$^{-1}$ for $\rm CH_3CN$, 
we increased the hidden neuron density $\alpha$ to 4 for $\rm H_2CO$ and 8 for $\rm CH_3CN$. We used a sequential optimization scheme that increases $N_s$ at each stage. There were four stages in total with step counts 500, 200, 100, and 100. After each stage, the number of selected configurations $N_s$ doubles. The initial $N_s$ was set at 256 for $\rm H_2CO$ and $256$ for $\rm CH_3CN$. The learning rate $\eta$ was set to 0.005 for both molecules. 
VSCF pretraining was used in all calculations. 

In Figure \ref{fig:h2co}, we present the absolute errors of five optimized energy levels compared to the vDMRG results for $\rm H_2CO$ as functions of $N_s$. We found that a moderate $N_s = 512$ was able to reduce the errors to less than 1 cm$^{-1}$ for all targeted states. The improvement in the error for the first three states is small for $N_s > 512$, while for states 3 and 4 the energy error dropped noticeably at $N_s=2048$. In Table \ref{tab:h2co_ch3cn}, we list the ZPE ($n=0$) and the four lowest vibrational transitions from the ground state ($n=1,2,3,4$) for both molecules. Due to the relatively uniform errors across all states and error cancellation effects, we found the transition energies for both molecules to be in good agreement with the vDMRG reference results ($\leq0.51$ cm$^{-1}$). It is also promising to see that,  
although the Hilbert space dimension increases quadratically going from $\rm H_2CO$ to $\rm CH_3CN$, 
doubling $N_s$ is sufficient to achieve similar accuracy for the larger molecule $\rm CH_3CN$. 

\begin{table}
    \centering
\begin{tabular}{p{2cm}p{1cm}p{2cm}p{2cm}p{2cm}}
    \hline    \hline
      & $n$ &  MBF &  DMRG & Error
    \\
    \hline
    $\rm H_2CO$ & 0 & 5773.69 & 5773.17 & +0.52
    \\
      & 1 & 1164.44 & 1164.25 & +0.19
    \\
      & 2 & 1245.58 & 1245.22 & +0.36
    \\
      & 3 & 1497.87 & 1498.31 & $-$0.44
    \\
      & 4 & 1744.20 & 1743.83 & +0.37
    \\ \hline
        $\rm CH_3CN$ & 0 & 9838.26 & 9837.41 & +0.85
    \\
      & 1 & 361.14 & 360.99 & +0.15
    \\
      & 2 & 361.14 & 360.99 & +0.15
    \\
      & 3 & 723.28 & 723.18 & +0.10
    \\
      & 4 & 723.69 & 723.18 & +0.51
    \\
    \hline    \hline
    \end{tabular}
    \caption{ZPE ($n=0$) and the four lowest vibrational transitions ($n=1,2,3,4$) for $\rm H_2CO$ (6 modes) and $\rm CH_3CN$ (12 modes) calculated with  MBF and compared to vDMRG reference results. The maximal $N_s$ for $\rm H_2CO$ and $\rm CH_3CN$ were 2048 and 4096, respectively. All energies are in cm$^{-1}$. 
    $N_{\rm max}=6$. The hidden neuron density was $\alpha = 4$ for $\rm H_2CO$ and $\alpha = 8$ for $\rm CH_3CN$.}
    \label{tab:h2co_ch3cn}
\end{table}

In this section, we understood that, while the performance of the MBF network is satisfactory for spectroscopic accuracy, it does not exceed the accuracy or computational efficiency achieved by TNSs. 
A recent benchmarking study showed that TTNSs 
deliver a ZPE and vibrational transitions for $\rm CH_3CN$ with error estimates below $10^{-3}$ cm$^{-1}$.
In terms of efficiency, the discrepancy between MBF and TNSs can be attributed to two issues. First, evaluating the energy of MBF is a global action. Although the selected-configuration scheme is more efficient in terms of the number of sampled states compared to the routinely used Monte Carlo method, it still requires collecting configurations in the total Fock space and evaluating their wavefunction amplitudes and local energies. By contrast, vDMRG reduces computational cost by optimizing the parameters of only one or two sites in each step.
Second, gradient-based energy minimizations can be prone to problems. Although MBF is much better at learning the vibrational energies than FNN, we still had to employ various techniques (such as pretraining and additional learning rate scheduling on top of the CoRe optimizer) 
to ensure stable optimization. Yet, the expressiveness of the MBF network could not fully be exploited by the current optimization scheme for the two larger molecules $\rm H_2CO$ and $\rm CH_3CN$. 
However, it should be emphasized that these issues in optimization are common to NQS methods and do not diminish the value of the modal backflow ansatz. Rather, the physically motivated MBF ansatz establishes a solid foundation for NQS-based vibrational structure calculations, positioning it to benefit from the ongoing advances in optimization techniques.

Another aspect for comparison is the number of free parameters: A vibrational MPS with bond dimension $m$ contains $\mathcal{O}(LN_{\rm max}m^2)$ parameters, while for the MBF ansatz with a single hidden layer, the number of free parameters scales as $\mathcal{O}(\alpha L^2 N_{\rm max})$. Although the scaling of the number of parameters for MBF is less favorable with respect to the number of modes $L$, it is unclear how $\alpha$ or $m$ would scale with $L$ for a given target accuracy. For example, for the 12-mode $\rm CH_3CN$ a fully-converged vDMRG calculation of various eigenstates requires a bond dimension around 100,\cite{baiardi2017vibrational} while the largest hidden layer density $\alpha$ we used in this work was 8, which makes the ratio between the number of parameters of the two approaches (ignoring the effect of prefactors) $m^2/(\alpha L)$ quite large. 
Still, for the systems we studied, increasing $\alpha$ beyond 8 led to only marginal improvements in energy, indicating that future advances in optimization methods will be crucial to fully exploit the expressiveness of the MBF ansatz and settle this comparison.

\section{Conclusions and Outlook}

Neural quantum states (NQSs) have emerged as a versatile ansatz for solving quantum many-body Hamiltonians. While there exist recent applications of NQS in electronic structure theory, implementations of NQS for the vibrational part of the molecular problem have so far been lacking.
In this work, we explored the feasibility of a tailored NQS to solve for the low-lying eigenstates of anharmonic vibrational problems. 

The centerpiece of our theory is the modal backflow (MBF) NQS design, which uses modal functions that depend on the input occupation number vectors to capture anharmonicity.
This tailored network significantly improved both the expressiveness and trainability compared to a conventional feedforward neural network (FNN) when applied for a Watson Hamiltonian.
We implemented a selected-configuration method for the calculation of expectation values and gradients, in place of the more commonly used Monte Carlo approach, to accommodate the highly peaked amplitude distributions of typical anharmonic eigenstates.

We incorporated a pretraining step using vibrational self-consistent field calculations, which can conveniently be carried out within the MBF framework. This pretraining step was found to be instrumental in promoting robustness of the optimization, especially when targeting excited states. First, we investigated the MBF ansatz with randomly sampled force constants to mimic different levels of anharmonicity and obtained accurate zero-point vibrational energies (ZPE) across all regimes of anharmonicity. Second, we applied the MBF ansatz in calculations with Watson Hamiltonians of three molecules, including the strongly anharmonic $\rm CH_3CN$. We were able to resolve both the ZPE and the low-lying vibrational transitions to spectroscopic accuracy. 

Our work extends the scope of NQSs to vibrational calculations, demonstrating their potential in quantum chemistry beyond the electronic problem. Using a comparison with FNNs, we showcased the clear advantages of the modal backflow formalism and underscored the merit of embedding the features of indistinguishable particles directly into the network architecture. 
Despite the universal approximation theorem, the approximating power of FNNs for anharmonic vibrational problems can be harnessed only with the inclusion of the essential MBF output layer.
Moreover, the effectiveness of using occupation number vector dependent modals revealed a distinctive structure of the correlation in anharmonic vibrational eigenstates, offering potentially transferable insights for other bosonic systems. 

We noted several aspects that present practical challenges for network optimization, such as the global nature of evaluating physical observables and limitations in current gradient-based optimization schemes. However, these challenges are not unique to the MBF ansatz, but rather general to the optimization of NQSs, an area of rapid development. Importantly, the physically motivated MBF ansatz provides the necessary theoretical groundwork that can be readily combined with future advances in optimization techniques, thus paving the way toward an accurate description of the vibrational structure of large molecular systems.

We envision several directions for future work. First, we can extend the modal basis from the harmonic eigenstates to general modal basis functions. The benefit of this is a potential reduction in the entanglement of the wavefunction and, hence, the number of important configurations to be selected. Second, the current paradigm of gradient-based optimization schemes struggles with the energy landscape of high-complexity networks. To fully exploit the expressiveness of NQS, more robust optimization schemes, such as second-order methods,\cite{drissi2024second} need to be explored. 
Finally, one can extend the current scope to include also pre-Born-Oppenheimer Hamiltonians, where fermionic and bosonic degrees of freedom of the molecule are treated on an equal footing (see, e.g., Ref. \cite{muolo2020nuclear}) by combining the MBF network with existing NQS ans\"atze for electrons.

\section*{Acknowledgement}
This project was supported by an ETH Postdoctoral Fellowship.

%

\end{document}